\documentclass[longbibliography,twocolumn,prl,aps,superscriptaddress,showpacs,amsmath,amssymb,floatfix]{revtex4-1}
\usepackage{graphicx}
\usepackage{amssymb}
\usepackage{amsmath}
\usepackage{epsfig}
\usepackage{color}
\usepackage{tabu}
\usepackage{mathtools}
\usepackage[colorlinks,linkcolor=blue,anchorcolor=blue,citecolor=blue,urlcolor=blue]{hyperref}
\usepackage{physics}
\usepackage{float}
\usepackage{diagbox}

\begin{document}

\title{Stochastic and long-distance level spacing statistics in many-body localization}
\author{Hong-Ze Xu}
\affiliation{CAS Key Laboratory of Quantum Information, University of Science and Technology of China, Hefei, 230026, China}
\author{Fei-Hong Liu}
\affiliation{CAS Key Laboratory of Quantum Information, University of Science and Technology of China, Hefei, 230026, China}
\author{Shun-Yao Zhang}
\affiliation{CAS Key Laboratory of Quantum Information, University of Science and Technology of China, Hefei, 230026, China}
\author{Guang-Can Guo}
\affiliation{CAS Key Laboratory of Quantum Information, University of Science and Technology of China, Hefei, 230026, China}
\affiliation{Synergetic Innovation Center of Quantum Information and Quantum Physics, University of Science and Technology of China, Hefei, Anhui 230026, China}
\affiliation{CAS Center For Excellence in Quantum Information and Quantum Physics}
\author{Ming Gong}
\affiliation{CAS Key Laboratory of Quantum Information, University of Science and Technology of China, Hefei, 230026, China}
\affiliation{Synergetic Innovation Center of Quantum Information and Quantum Physics, University of Science and Technology of China, Hefei, Anhui 230026, China}
\affiliation{CAS Center For Excellence in Quantum Information and Quantum Physics}
\date{\today }

\begin{abstract}
From random matrix theory all the energy levels should be strongly correlated due to the presence of all off-diagonal entries.
In this work we introduce two new statistics to more accurately characterize these long-distance interactions in the disordered many-body systems 
with only short-range interaction. In the $(p, q)$ statistics, we directly measure the long distance energy level spacings, while in the
second approach, we randomly eliminate some of the energy levels, and then measure the reserved $\eta\%$ energy levels using nearest-neighbor 
level spacings. We benchmark these results using the results in standard Gaussian ensembles. Some analytical distribution functions with extremely
high accuracy are derived, which automatically satisfy the inverse relation and duality relation. These two measurements satisfy the same universal scaling law during the 
transition from the Gaussian ensembles to the Poisson ensemble, with critical disorder strength and corresponding exponent are independent of these
measurements. These results shade new insight into the stability of many-body localized phase and their universal properties in the disordered many-body 
systems. 
\end{abstract}

\maketitle

Sixty years ago Anderson demonstrated that a single particle can be localized by a random potential from destructive interference, known as Anderson Localization (AL) 
\cite{anderson1958absence,abrahams201050, abrahams1979scaling}.  Meanwhile Wigner developed the idea of random matrix theory (RMT) to describe the energy level spacing 
in heavy atom unclear with strong interaction \cite{wigner1951statistical, mehta2004random, wigner1993characteristicI, wigner1993characteristicII,dyson1962statistical}.
These two theories correspond to two different physics. In AL, the energy levels are spatially localized with energy level spacings described by Poisson distribution.
In RMT, however, the wave functions are spatially delocalized with strong repulsive interaction, and the level spacings are described by Wigner surmise in 
Gaussian ensembles \cite{altshuler1986repulsion,beenakker2015random}. The combination of disorder and interaction can give rise to 
many-body localization (MBL) \cite{pal2010many,vznidarivc2008many,kjall2014many,altman2018many}, which is an important concept that has been 
intensively explored in recent years. The transition from ergodic phase to the MBL phase is observed by increasing disorder strength \cite{oganesyan2007localization,vosk2015theory,luitz2015many,serbyn2016spectral,zhang2018universal}. 
In ergodic phase, the eigenstate thermalization hypothesis (ETH) is valid \cite{tasaki1998quantum,rigol2008thermalization}, with entanglement entropy in accord with the 
volume law \cite{vitagliano2010volume,khemani2017two} and the level spacings follow the Wigner surmise \cite{canovi2011quantum}. 
Conversely, the  MBL phase breaks the ETH \cite{ponte2015many}, violates the volume law \cite{devakul2015early,huse2014phenomenology}, with level spacings 
follow the Possion law \cite{geraedts2016many}. More intriguing features of the MBL phase can be found in \cite{nandkishore2015many,alet2018many,abanin2018ergodicity}. 

\begin{figure} 
\centering
\includegraphics[width=0.4\textwidth]{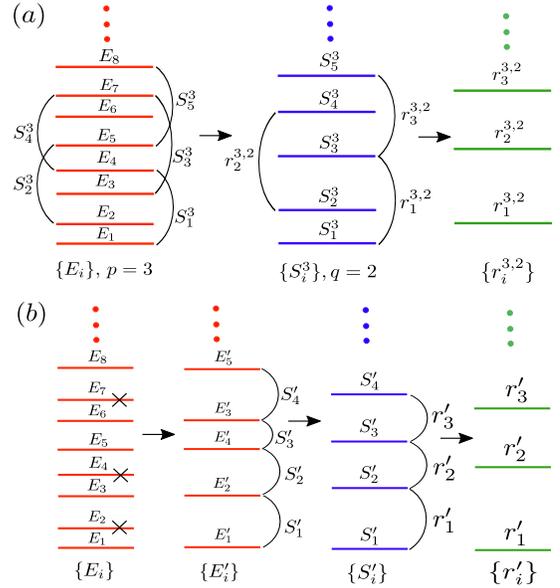}
\caption{Two new statistics developed in this work. (a) $(p,q)$ level spacing ratios and (b) stochastic 
level spacing ratios.}
\label{fig-fig1}
\end{figure}

The disordered many-body models with short range interaction are obviously totally different from that in RMT, in which
all the matrix entries with identical independent distribution are presented. This long range feature can not be captured
by the $r$-statistics based on nearest-neighbor level spacings \cite{atas2013distribution,chavda2013probability,janarek2018discrete,oganesyan2007localization}.
In this work we present two new approaches to characterize the long-distance level spacings in a disordered many-body models. 
In the first approach, we introduce the $(p,q)$ statistics to measure the level spacing ratios between distant energy levels. Meanwhile, we randomly 
eliminate some of the energy levels, keeping only $\eta\%$ of them, and then measure the reserved levels using $r$-statistics. 
Some analytical expressions are obtained for these statistics, which automatically satisfy the inverse relation and duality relation.
We find that the results in the disordered many-body models agree well with the predictions from the standard Gaussian orthogonal ensemble (GOE) and 
unitary ensemble (GUE). These new statistics also satisfy the universal scaling laws during phase transition from the Gaussian ensembles to the 
Poisson ensemble (PE), with critical disorder strength and its exponent independent of these measurements. Our results shade new insight to the 
strong energy level interactions and their stability in disordered many-body systems.

\begin{table}
\centering
\caption{Validity of the unified distribution for the three ensembles. $\langle r\rangle_{\text{RM}}$ is obtained from Gaussian random matrices with size $N = 10^4$ averaged over 
$2\cdot 10^4$ realizations, and $\langle r\rangle_{\text{th}}$ is obtained from Eq. \ref{eq6}.}
\begin{tabular}{p{0.45cm}<{\centering} p{0.45cm}<{\centering} p{1.1cm}<{\centering}p{1.1cm}<{\centering}p{1.1cm}<{\centering}p{1.1cm}<{\centering}p{1.1cm}<{\centering}p{1.1cm}<{\centering}}    
\toprule
\multicolumn{1}{c}{ } &   & \multicolumn{2}{c}{GOE ($\beta = 1$)} & \multicolumn{2}{c}{GUE ($\beta = 2$)} & \multicolumn{2}{c}{GSE ($\beta = 4$)} \\ \hline 
p &q & $\langle r\rangle_{\text{RM}}$ &$\langle r\rangle_{\text{th}}$  & $\langle r\rangle_{\text{RM}}$ &$\langle r\rangle_{\text{th}}$  & $\langle r\rangle_{\text{RM}}$ &$\langle r\rangle_{\text{th}}$ \\ 
\hline
1 & 1 &  0.5307 & 0.5359 & 0.5997 & 0.6027 &  0.6744 & 0.6762\\
1 & 2 &  0.5548 & 0.5505 & 0.6266 & 0.6185 &  0.7006 & 0.6919\\
1 & 3 &  0.5606 & 0.5585 & 0.6316 & 0.6271 &  0.7046 & 0.7002\\
1 & 4 &  0.5626 & 0.5635 & 0.6333 & 0.6324 &  0.7059 & 0.7054\\
2 & 1 &  0.7396 & 0.7383 & 0.7870 & 0.7866 &  0.8334 & 0.8340\\
2 & 2 &  0.6744 & 0.6762 & 0.7336 & 0.7335 &  0.7908 & 0.7902\\
2 & 3 &  0.6940 & 0.6919 & 0.7539 & 0.7480 &  0.8090 & 0.8027\\
2 & 4 &  0.7000 & 0.7002 & 0.7589 & 0.7556 &  0.8128 & 0.8092\\
3 & 1 &  0.8202 & 0.8198 & 0.8541 & 0.8540 &  0.8867 & 0.8865\\
3 & 2 &  0.7802 & 0.7811 & 0.8258 & 0.8256 &  0.8665 & 0.8669\\
3 & 3 &  0.7460 & 0.7464 & 0.7970 & 0.7964 &  0.8432 & 0.8429\\
3 & 4 &  0.7618 & 0.7606 & 0.8127 & 0.8087 &  0.8568 & 0.8530\\
4 & 1 &  0.8624 & 0.8624 & 0.8888 & 0.8887 &  0.9139 & 0.9138\\
4 & 2 &  0.8329 & 0.8335 & 0.8685 & 0.8686 &  0.8998 & 0.9000\\
4 & 3 &  0.8132 & 0.8144 & 0.8545 & 0.8544 &  0.8896 & 0.8897\\
4 & 4 &  0.7909 & 0.7902 & 0.8347 & 0.8344 &  0.8736 & 0.8740\\
10& 3 &  0.9239 & 0.9242 & 0.9412 & 0.9412 &  0.9557 & 0.9557\\
10& 5 &  0.9157 & 0.9156 & 0.9355 & 0.9356 &  0.9515 & 0.9515\\
10& 8 &  0.9057 & 0.9054 & 0.9285 & 0.9284 &  0.9462 & 0.9459\\
10&10&  0.8935 & 0.8947 & 0.9184 & 0.9209 &  0.9382 & 0.9419\\
10&11&  0.8997 & 0.9018 & 0.9240 & 0.9264 &  0.9426 & 0.9461\\
10&12&  0.9022 & 0.9055 & 0.9259 & 0.9292 &  0.9438 & 0.9481\\
\toprule
\end{tabular}
\label{tableI}
\end{table}

{\it Physical Models and Measurements}. We consider the following disordered spin-${1 \over 2}$ spin chain \cite{vznidarivc2008many,serbyn2014quantum,serbyn2014interferometric,roy2018dynamical}
\begin{equation}
	H = \sum_{i=1}^L J(e^{i\theta}S_i^+S_{i+1}^{-}+ \text{h.c.}) + J_z S_i^zS_{i+1}^{z} + h_i S_i^z,
\label{model}
\end{equation}
where $h_i$ is a uniform random potential in $[-W,W]$, $S_{i+L} = S_i$ from periodic boundary condition and $L$ is the total length of chain. 
In case $J_z= J$, previous inversions have unveiled a critical disorder strength $W_c \sim 3 - 4$ \cite{pal2010many,luitz2015many,serbyn2015criterion,
luitz2016extended}. When $W < W_c$, it gives an ergodic phase with energy level spacings to be described by GOE ($\theta = 0$) or GUE ($\theta \ne 2\pi n/L$, $n\in \mathbb{Z}$ \cite{pbc}). 
In contrast, in the MBL phase ($W>W_c$), it follows the Poisson distribution. In the above model, the total spin $S_{\text{tot}}^z = \sum_i S_i^z$ 
is conserved and in the numerical simulation, we will focus on the largest subspace with $S_\text{tot} = L/2 - [L/2]$. 
Let us denote the energy levels to be $\{E_i\}$ sorted in ascending order, and define the $p$-level spacing as $S_i^p = E_{i+p}-E_{i}$, then we can define the 
$(p, q)$ level spacing ratios as 
\begin{equation}
r = r_i^{p,q} = \frac{\min (S^{p}_{i+q},S^{p}_i)}{\max (S^{p}_{i+q},S^{p}_i)}, \quad i,p,q=1,2,\cdots.
\label{eq-ri}
\end{equation}
The picture for this definition is shown in Fig. \ref{fig-fig1} (a). When $p < q$, there is no overlap between the two $p$-level spacings; while when $p > q$,
some overlap between them exists. In the second approach, we randomly eliminate some of the energy levels, keeping only $\eta\%$ of them, which is then measured using 
the nearest-neighbor $r$-statistics (see Fig. \ref{fig-fig1} (b)). These two measurements aim to explore the long-distance energy level interactions induced 
by the random potential, which is a typical feature RMT. 

{\it Random Gaussian Ensembles}. To further benchmark these results, we consider these measurements in the Gaussian ensembles belonging to GOE, GUE and Gaussian 
symplectic ensemble (GSE) for $\beta =$ 1, 2, 4, respectively. The joint distribution function (JDF) is \cite{mehta2004random}, 
\begin{equation}
\rho_{\beta}(E_1,...,E_N) = C_{\beta,N} \prod_{i<j} |E_i-E_j|^{\beta}e^{-\frac{\beta}{2}\sum_i E_i^2}.
\end{equation}
To determine the distribution of $r_i^{p,q}$, we need to consider the JDF at least with order $N = p+q+1$. In the above equation, we 
can define $E_1=\lambda_0$, $E_i = \lambda_0 +\sum_{j=1}^{i-1} \lambda_j$ $(i=2, 3, \ldots,N)$, then we obtain
\begin{equation}
P_{\beta}(r,p,q)  \varpropto \int \prod_{n=0}^{p+q} \text{d}\lambda_n \rho_{\beta}(\Lambda_0^0,\ldots,\Lambda_0^{p+q})
\delta(r-\frac{\Lambda_{1+q}^{p+q}}{\Lambda_1^p}),
 \label{eq-Pr}
\end{equation}
where $\Lambda_a^b = \sum_{i=a}^b \lambda_i$. For $p=q=1$ and $N=3$, 
the distribution $P_{\beta}(r,1,1)$ is studied in \cite{atas2013distribution}. For $N = 4$, we can obtain $P_{\beta}(r,1,2)$ and $P_{\beta}(r,2,1)$, which can 
also be found in \cite{atas2013joint}. For $N> 4$, this expression will become formidable. By analysing the solvable cases with $N \le 4$, we find that 
the following quadratic function will always be presented in square root, 
\begin{equation}
	g(r,k) = a_k + \text{sgn}(q-p) kr +a_k r^2,
\label{eq5}
\end{equation}
where $k=|p-q|+1$ and $a_k = a_{k-1}+k$ with $a_0 = 0$, and $\text{sgn}(x)$ is the sign function with $\text{sgn}(0)=1$. Here $k$ can be regarded as 
how many energy levels are spaced between two spacings for $p < q$, or the overlap between them for $p > q$. Obviously, $g(k, r)> 0$ for any $r$ and 
$k$. The key observation is that we assert this quadratic function is essential for the distribution function.

For $p\leq q$, we obtain a very accurate approximated distribution $P_{\beta}(r,p,q)$, which is given by
\begin{equation}
P_{\beta}(r,p,q) = C_{\beta}\frac{(r+r^2)^{\beta_p}}{[g(r,k)]^{1+\frac{3}{2}\beta_p}},
\label{eq6}
\end{equation}
where $\beta_p = \frac{p(p+1)}{2}\beta + p-1$ and $C_{\beta}$ is the normalized constant. This unified distribution for the three ensembles
are one important finding in this work. It has a number of interesting features. (I) It automatically satisfies the inverse relation \cite{atas2013distribution}, 
\begin{equation}
P_{\beta}(r,p,q) = P_{\beta}(1/r,p,q)/r^2,
\label{eq-dual}
\end{equation}
as required by definition of Eq. \ref{eq-ri}. (II) When $p = q$, we find $k = 1$ and $a_k = 1$, which will yields duality between GOE and GSE as following, 
\begin{equation}
	P_1(r, 2p, 2p) = P_{4}(r, p, p), \quad p=1, 2, 3, \cdots.
\end{equation}
It means that the $(2p, 2p)$ distribution in GOE is the same as ($p, p$) distribution in GSE. Two examples ($p=1, 2$) for this duality can be found in Table 
\ref{tableI},  and more examples are listed in Ref. \cite{ppdata}. 
The duality for $p = 1$ has been unveiled by Forrester in \cite{forrester2009random} from the exact duality in sense of JDF. Our numerical results may 
suggest more relations in these ensembles. For instance, we may even find $P_1(r, 7, 7) = P_2(r, 5, 5)$ with $\langle r\rangle_\text{RM} = 0.8602$ \cite{ppdata}. 

For $p>q$, there is an overlap between adjacent $p$-level spacings. We find that the distribution of $r$ can also be given by Eq. \ref{eq6} with some different $\beta_p$, which 
can be written as $\beta_p = a p^2 +b p +c$ for the same $q$ and $\beta$. The values for $a$, $b$ and $c$ are  fitting parameters \cite{abcdata}. 
In Table \ref{tableI} we present $\langle r \rangle_{\text{th}}$ form  Eq. \ref{eq6} and $\langle r\rangle_{\text{RM}}$ from the Gaussian ensembles,
which exhibit excellent agreement between these two descriptions. 

\begin{figure}
\centering
\includegraphics[width=0.45\textwidth]{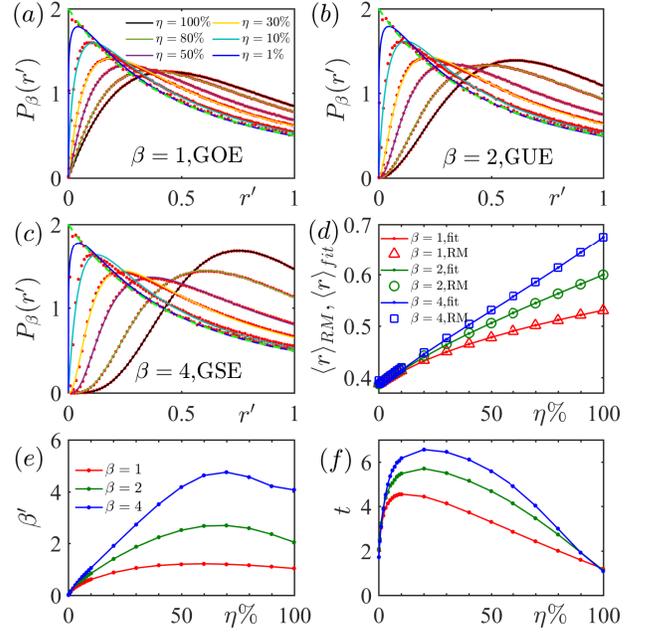}
\caption{(a) - (c) Distribution of the stochastic level spacing ratios for the three Gaussian ensembles. 
Red dots are the numerical results from Gaussian matrices with size $N = 10^4$ averaged over $2\cdot 10^4$ realizations, and the solid lines are corresponding 
best fit with Eq. \ref{eq9}. The green dashed lines are given by Eq. \ref{eq-PD}. (d) The average value of $r^\prime$, where $\langle r^\prime \rangle_{\text{RM}}$ (open symbols) 
are determined numerically from Gaussian ensembles and $\langle r^\prime \rangle_{\text{fit}}$ (solid lines) are calculated from Eq. \ref{eq9}. 
(e) - (f) fitted parameters of $\beta^\prime$ and $t$ as a function of $\eta$.}
\label{fig-fig2}
\end{figure}

Next we discuss the stochastic level spacing ratios $r^\prime$ by considering some randomly selected energy levels (see Fig. \ref{fig-fig1} (b)). We find that the
distribution in these three ensembles can still be written as,
\begin{equation}
P_{\beta}(r^\prime) =C^\prime_{\beta} \frac{(r^\prime + r^{\prime^2})^{\beta^\prime}}{(1+t r^\prime + r^{\prime^2})^{1+\frac{3}{2} \beta^\prime}},
\label{eq9}
\end{equation}
where $C^\prime_{\beta}$ is a normalized constant and $t$ and $\beta^\prime$ are fitted parameters, which depends on $\eta$. This new distribution also satisfies
the inversion symmetry in Eq. \ref{eq-dual}. The numerical results and their best fitting can be found in Fig. \ref{fig-fig2} (a) - (d), and their corresponding
$\beta'$ and $t$ are presented in Fig. \ref{fig-fig2} (e) - (f). When $\eta \rightarrow 0$, we find $\beta' \rightarrow 0$ and $t\rightarrow 2$, which realizes a
transition from Gaussian ensembles to Poisson distribution,
\begin{equation}
	P_\text{PE}(r^\prime)={2 \over (1+r^\prime)^2}.
	\label{eq-PD}
\end{equation}
This is expected since the stochastic energy levels in the small $\eta$ limit has completely erased the correlation between them, which is the essential assumption 
during the derivation of Poisson distribution \cite{mehta2004random}. 
With this method we may realize some distributions in Laguerre $\beta$-ensembles \cite{dumitriu2002matrix,dumitriu2006global}, which exhibit some dualities
in JDP \cite{forrester2009random}. These results also demonstrate the excellent predictability of Eqs. \ref{eq6} and \ref{eq9}. 

{\it Poisson Ensemble}. Now we consider the $(p, q)$ statistics for Poisson random variables $\{E_i\}$. Using the previous method, we find
\begin{equation}
P_{\text{PE}}(r,p,q) = \int_0^\infty \prod_{i=0}^{p+q} \text{d}\lambda_i  e^{-\Lambda_0^{p+q}} \delta(r-\frac{\Lambda_{1+q}^{p+q}}{\Lambda_1^p}). 
 \label{eq11}
\end{equation}
For $p\leq q$, the distribution $P_{\text{PE}}(r,p,q)$ is given by
\begin{equation}
P_{\text{PE}}(r,p,q) = \frac{1}{C_p}  \frac{r^{p-1}}{(1+r)^{2p}}.
\label{eq12}
\end{equation}
When $p =1$, this function yields Eq. \ref{eq-PD}. Then we obtain $\langle r\rangle_{\text{PE}} = _2F_1(2p,1+p;2+p;-1)\Gamma(1+p)p/_2F_1(p,2p;1+p;-1)\Gamma(2+p)$,
where $_2F_1$ is the hypergeometric series \cite{gasper2004basic,bailey1928products}. The values of
this mean value for $p = 1$ to $6$ can be found in Ref. \cite{ppdata}. This distribution also satisfies the inverse relation
in Eq. \ref{eq-dual}. 

For $p>q$, we denote $x=\Lambda_{q+1}^{p}$, then Eq. \ref{eq11} can be rewritten as 
\begin{align}
P_{\text{PE}}(r,p,q) = &\int_0^\infty  \text{d}x  \prod_{i=1}^{q} \text{d}\lambda_i   \prod_{j=p+1}^{p+q} \text{d}\lambda_j  
e^{-\Lambda_1^q- \Lambda_{p+1}^{p+q}}  \nonumber \\
&\cdot \frac{x^{p-q-1}}{(p-q-1)!e^{-x}} \delta(r-\frac{x+\Lambda_{p+1}^{p+q}}{\Lambda_1^q + x}),
\label{eq13}
\end{align}
\begin{figure}
\centering
\includegraphics[width=0.45\textwidth]{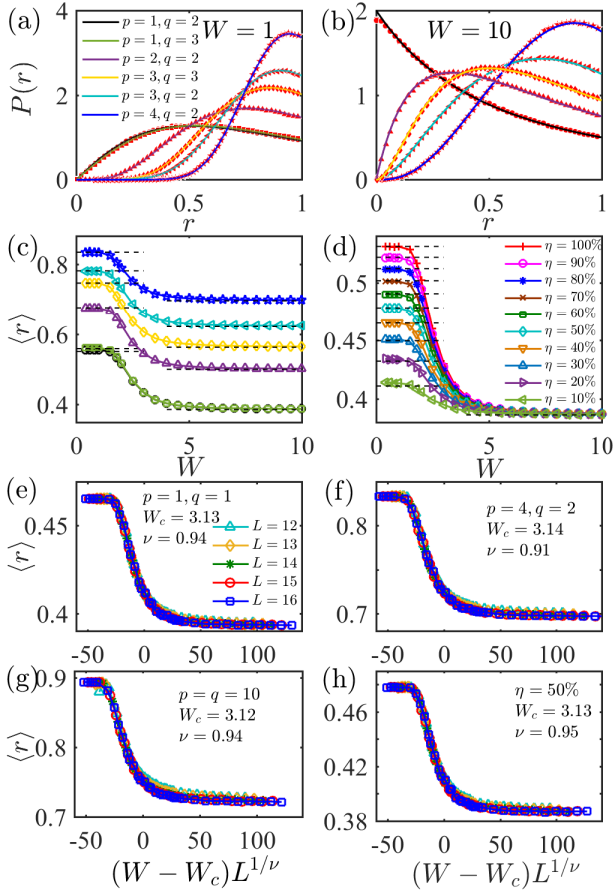}
\caption{GOE with $\theta=0$. (a) In ergodic phase ($W=1$) and (b) in MBL phase ($W=10$), where symbols are data averaged over 4000 realizations in a chain with $L=14$, and
dashed lines are these given by Eqs. \ref{eq6}, \ref{eq11} and \ref{eq14}. (c) The $(p,q)$ statistics as a function of $W$.  The symbols are the same as the 
six cases studied in (a). (d) The stochastic statistics for different reserved rate $\eta$. In (c) and (d) the horizontal dashed lines are given by $\langle r\rangle_{\text{th}}$, 
$\langle r^\prime\rangle_{\text{fit}}$ or $\langle r\rangle_{\text{PE}}$. (e) - (h) The data collapse used to extract the critical disorder strength $W_c$ and exponent $\nu$ for the 
different measurements. These data are averaged over 7000, 5000, 4000, 1000 and 300 realizations for $L=12-16$, respectively.}
\label{fig-fig3}
\end{figure}
\begin{figure}
\centering
\includegraphics[width=0.45\textwidth]{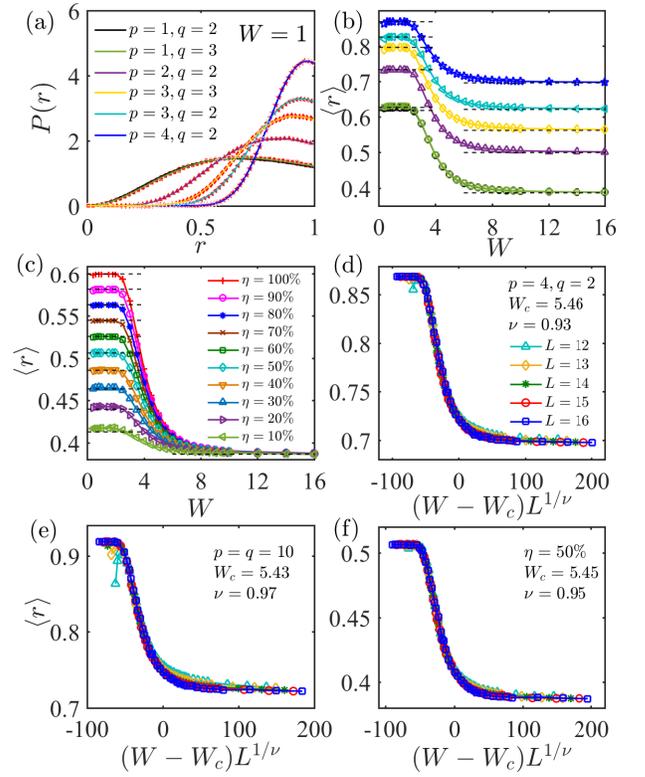}
	\caption{GUE with $\theta=\pi/28$. (a) In ergodic phase ($W=1$). (b) - (c) The variation of $\langle r\rangle$ for different $(p,q)$ and $\eta$, respectively. (e) - (f) 
	Methods to extract $W_c$ and $\nu$. The other legend descriptions are the same as that in Fig. \ref{fig-fig3}.}
\label{fig-fig4}
\end{figure}
For $0<r<1$, we find $P_{\text{PE}}(r,p,q)$ can be expressed as
\begin{equation}
P_{\text{PE}}(r,p,q) =  \frac{2r^{p-1}}{(1+r)^{2q}} \sum_{m=0}^{2q-1} r^m w_m,
\label{eq14}
\end{equation} 
where $w_m$ is given by
\begin{equation}
w_m = (-1)^n \mathcal{Y}_{q,n+1} f_{m-n,q-1}^p f_{-q,m-n-(q+1)}^p,
\label{eq15}
\end{equation}
where $n=[m/2]$ and $\mathcal{Y}_{i,j}$ is a number of the $i$-$\text{th}$ row and $j$-$\text{th}$ column of the Yang Hui's triangle \cite{yadav2011ancient}, 
and $f_{a,b}^p= \Gamma(p+b+1)/\Gamma(p+a)$. For $q = 1$, Eq. \ref{eq14} will yields,
 \begin{equation}
 P_{\text{PE}}(r,p,1) =
  \left\{
\begin{aligned}
    & { r^{p-1} (p + (p-1)r) \over (1 +r)^2}, \quad 0<r<1 \\
   & \frac{p-1+p r}{r^p (1+r)^2},  \quad r>1
\end{aligned} 
\right.
 \end{equation}
which was also shown in \cite{atas2013joint}. These expressions also satisfy Eq. \ref{eq-dual}.

{\it Many-body systems and MBL}.  Finally we use the above results to understand the distribution in Eq. \ref{model}. 
In GOE (Fig. \ref{fig-fig3}), we consider $J=1/2$, $J_z=1$ and $\theta =0$ and in GUE (Fig. \ref{fig-fig4}), we consider $J=J_z=1$ and $\theta = \pi/28$.
We employ the exact diagonalization (ED) method to study these two new statistics in a finite system and compare them with the analytical expressions obtained in 
the previous paragraphs. We normalize the eigenvalues using
$\varepsilon = (E-E_{\min})/(E_{\max}-E_{\min})$, where $E_{\min}$ $(E_{\max})$ are the energies of the ground state (highest excited state) of the system. 
For different $\varepsilon$, there is a different $W_c$, indicating of many-body mobility edges \cite{mondragon2015many,luitz2015many,laumann2014many,baygan2015many}. In this paper, we mainly discuss the case of $\varepsilon= 0.5\pm 0.15$ in the middle of the spectra. 
For $(p,q)$ level spacing ratios, the results are shown in Fig.  \ref{fig-fig3} (a) - (b) and Fig. \ref{fig-fig4} (a), with $W = 1$ to the physics in GOE or GUE,
and $W=10$ to that in PE. In Fig. \ref{fig-fig3} (c) - (d) and Fig. \ref{fig-fig4} (b) - (c), we give the variation of $\langle r\rangle$ with $W$ for different $(p,q)$ and $\eta$.  
In Fig. \ref{fig-fig3} (e) - (h) and Fig. \ref{fig-fig4} (d) - (f), we show our new statistics also satisfy the scaling laws. For $\theta = 0$ (GOE), we obtain
$W_c\approx 3.13\pm 0.02$ and $\nu\approx 0.94\pm0.03$, in consistent with Ref. \cite{luitz2015many} with $W_c\approx 3.72$ and $\nu\approx 0.91 \pm 0.03$ \cite{Wcdiff}. For $\theta = \pi/28$ (GUE), we obtain $W_c\approx 5.44\pm0.03$ and $\nu\approx 0.95\pm0.04$. All these results shown that the critical disorder strength $W_c$ and the exponent $\nu$ do not change significantly in these two measurements with different $(p, q)$ and $\eta$, demonstrating their universality in Gaussian ensembles. 

{\it Conclusion}. We introduce two new approaches to characterize the long-distance energy level interactions in the disordered many-body systems. 
Some analytical distributions with high accuracy are obtained by benchmarking these results against the RMT. These expressions also automatically 
satisfy the inverse relation and duality relation. These new statistics also yield some universal scaling laws, in which the critical exponent and
critical disorder strength are almost independent of the choice of the two different statistics. Our results indicate that although the physical 
models are made by short-range interaction, their energy levels are long-range correlated. These features may also be revealed from the 
entanglement entropy  \cite{ponte2015many,vasseur2015quantum,bera2015many,luitz2016extended}, inverse participation ratios  \cite{serbyn2013local,iyer2013many,torres2015dynamics} and spin imbalance \cite{luitz2016extended,zhang2016floquet} using only a fraction of spectra. 
Our results demonstrate the robustness of MBL phases and their universal features. 

\textit{Acknowledgements.} We thank Prof. Dang-Zheng Liu for valuable discussion. This work is supported by the National Youth Thousand Talents Program (No. KJ2030000001), the USTC start-up funding (No. KY2030000053), the NSFC (No. 11774328) and the  National Key Research and Development Program of China (No. 2016YFA0301700).

\bibliography{ref}

\begin{thebibliography}{59}
\bibitem [{pbc()}]{pbc}%
  \BibitemOpen
  \href@noop {} {}\bibinfo {note} {We consider periodic boundary condition, in
  which when $\theta = 2 \pi n / L$, the phases between the neighboring sites
  can be gauged out by defining $\widetilde{S}_k = e^{i(k-1) \theta}S_k$, in
  which we find $\widetilde{S}_1^-= e^{iL\theta}S_1^- = S_1^-$. In a system
  with open boundary condition, this phase will not yields new different
  observations in the statistics of energy levels.}\BibitemShut {Stop}%
\bibitem [{ppd()}]{ppdata}%
  \BibitemOpen
  \href@noop {} {}\bibinfo {note} {We find that $\langle r\rangle_\text{RM}$ in
  both GOE and GSE are 0.8434, 0.8739, 0.8940, 0.9083 for $p=q = $ 6, 8, 10,
  12, respectively. We also find this value is 0.8602 in GOE ensemble for $p =
  7$, while in GUE, it is 0.8605 for $p = 5$. The mean value $\langle
  r\rangle_\text{PE}$ are 0.3863, 0.500, 0.5625, 0.6042, 0.6348, 0.6586 for $p
  =q=1$ to $6$, respectively.}\BibitemShut {Stop}%
\bibitem [{abc()}]{abcdata}%
  \BibitemOpen
  \href@noop {} {}\bibinfo {note} {$(a,b,c)$ for $\beta=1$, 2, 4 are:
  $(1.43,-1.8,0.7)$, $(2.22,-2.56,1)$, $(3.83,-4.47,2.3)$ for $q=1$;
  $(0.98,-1.5,0.15)$, $(1.62,-2.2,-0.25)$, $(2.84,-3.4,-1)$ for $q=2$;
  $(0.83,-1.4,-1)$, $(1.37,-1.7,-3.4)$, $(2.42,-2.48,-7)$ for $q=3$;
  $(0.6,0.31,-11.2)$, $(1.06,0.56,-19.4)$, $(1.81,2.52,-39.4)$ for $q=5$;
  $(0.37,4.46,-46.8)$, $(0.607,9.1,-88.6)$, $(0.87,20.3,-176)$ for
  $q=8$.}\BibitemShut {Stop}%
\end{thebibliography}

\end{document}